# Liquidity Constraints and Demand for Healthcare: Evidence from Danish Welfare Recipients[1]

Frederik Plesner Lyngse[2]

*University of Copenhagen*

October 2020

## Abstract

Are low-income individuals relying on government transfers liquidity constrained by the end of the month to a degree that they postpone medical treatment? I investigate this question using Danish administrative data comprising the universe of welfare recipients and the filling of all prescription drugs. I find that on transfer income payday, recipients have a 52% increase in the propensity to fill a prescription. By separating prophylaxis drugs used to treat chronic conditions, where the patient can anticipate the need to fill the prescription, e.g. cholesterol-lowering statins, I find an increase of up to 99% increase on payday. Even for drugs used to treat acute conditions, where timely treatment is essential, I find a 22% increase on payday for antibiotics and a 5-8% decrease in the four days preceding payday. Lastly, exploiting the difference in day the doctor write the prescription and the day the patient fill it, I show that liquidity constraints is the key operating mechanism for postponing antibiotic treatment.

[1] I thank Torben Heien Nielsen, Søren Leth-Petersen, Nanna Skovgaard and Itzik Fadlon, for input and support at all stages in this project. I also thank Jeffrey Clemens, Matthias Sutter, Gordon Dahl, Janus Laust Thomas, Maja Skov Paulsen, as well as participants at the UCPH PhD Seminar 2017; The Danish Ministry 2017; DGPE 2017 conference; Behavior and Incentives 2017 course; UCSD Applied Economics Lunch 2018; DAEINA 2018 conference; ESPE 2018 conference; VIVE workshop on Public Policies, Health, and Health Behaviors 2018; EDGE 2018 conference; UCPH workshop on Behavioral responses to health innovations and the consequences for socioeconomic outcomes 2019; UCPH workshop on Behavioral Insights for Healthy Ageing 2019; for helpful discussions and comments. I thank Ida Lind for inspiration for this project.
[2] Department of Economics & Center for Economic Behavior and Inequality, University of Copenhagen & Danish Ministry of Health, fpl@econ.ku.dk.

# 1. Introduction

Patient adherence to medical treatment is a primary determinant of treatment success. The World Health Organization (WHO) defines poor adherence to treatment of as a worldwide problem of striking magnitude (WHO 2003). Estimates from the U.S. find that 50% of medications are not taken as prescribed. This lack of adherence is estimated as causing approximately 120,000 American deaths, at least 10% of hospitalizations, and costing the American healthcare system USD 100-289 billion a year (4-11% of total healthcare costs) (Viswanathan et al. 2012).

Understanding why patients do not follow medical advice is a key question for developing policies to improve patient adherence. Economic constraints arguably constitute an important factor for non-adherence. Acknowledging this, most countries have some kind of health insurance, which often include out-of-pocket costs for the patient. However, these out-of-pocket costs may cause individuals living paycheck-to-paycheck to postpone necessary treatment—especially towards the end of the month. Low-income individuals relying on social insurance is a group of individuals that are likely financial constrained—also with respect to purchasing necessary medical treatment. In fact, surveys suggest that 42% of welfare recipients have refrained from buying prescription drugs for financial reasons (Trygfonden 2014). This is also a group, where the marginal benefit of additional insurance is relatively high.

In this paper, I address this question by studying prime-age individuals living off social insurance and their behavior of purchasing prescription drugs around transfer income payday. Specifically, I study Danish individuals living off welfare benefits, which is the social insurance for individuals who cannot support themselves and their family in any other way. In 2005, this social insurance affected 6% of prime-age individuals (Rosdahl & Nærvig Petersen 2006). As welfare benefits are means- and income-tested, recipients are not allowed to have any other income nor have any savings exceeding DKK 10,000 (≈USD 1,600). This makes recipients are comparable in income and wealth.

I document an increased propensity to fill prescriptions on payday and investigate underlying mechanisms. To do so, I leverage a large panel of Danish administrative data with rich details on all prescription drug purchases and all government transfers. The data covers the entire Danish population of welfare benefit recipients 2000-2016. An important feature of the data is the ability to link government transfers to daily measures of medical spending on prescription drugs at the individual level. This allows me to pinpoint variation in drug purchases around the transfer income payday. Moreover, the rich information on the prescription drug types allows me to separate drugs into different categories, according to their Anatomical Therapeutic Classification (ATC) code. This is essential for distinguishing stockpiling of drugs used to treat chronic conditions from drugs used to treat acute conditions, and hence, understand underlying mechanisms. Furthermore, the data allows me to identify patient groups, for whom the cost of postponing acute treatment is especially high and investigate spillovers to the next generation. Lastly, by linking prescriptions to the last doctor visit, I proxy for the day of prescription. This allows me to calculate the number of days the patient postpone filling the prescription. In doing so, I can separate the choice of seeking medical attention from the choice of purchasing the prescribed treatment. I estimate a non-parametric monthly event-study regression at the daily level on the propensity to purchase a prescription drug for all individuals on welfare benefits.

I contribute with four novel findings. First, I group all types of drugs and show that on payday there is a 52% increase in the propensity to fill a prescription relative to an average day with a baseline of 1.3%. Furthermore, I find a decrease of up to 8% in the four days preceding payday, indicating that individuals postpone prescription drug purchases to the transfer income payday. Second, I separate prophylaxis drugs used to treat chronic conditions that patients can stockpile, e.g. cholesterol-lowering statins. For these types of drugs, I find an increase in the propensity to fill a prescription of up to 99% on payday, indicating that stockpiling is a likely mechanism for these types of drugs.

Third, I separate drugs used to treat acute conditions. In particular, I study antibiotics, used to treat bacterial infections, for which the infection is assumed uncorrelated with payday, but where immediate treatment is crucial for successful treatment. For these types of drugs, I find an increase of 22% at payday and a decrease in the four days preceding payday. Next, I investigate the underlying mechanisms for this behavior. I show that on payday there is a 9% increase in doctor visits for individuals on welfare benefits, suggesting that some patients may postpone the free doctor visit due to the expectation of the subsequent out-of-pocket cost of the prescribed treatment. Lastly, I show that liquidity constraints is a key operating mechanism for postponing antibiotic treatment, as patients going to the doctor one day before payday postpone treatment exactly one day, patients going to the doctor two days before payday postpone two days, etc. up to eight days before payday, whereafter there is no effect.

Fourth, I separate vulnerable patients, for whom the cost of postponing antibiotic treatment is especially high, e.g. pregnant women and children. Even for pregnant women, I find a similar behavior around payday.

My results show that low-income individuals living off social insurance postpone necessary treatment, which should be treated promptly in order not to lead to complications, such as the bacterial infection spreading (Longo et al. 2013; Mayo Clinic 2019b). Thus liquidity constraints for individuals living off government transfers translate into inequalities in access to timely treatment and potentially real health effects. Furthermore, I find a similar behavior of pregnant women postponing antibiotic treatment, which may increase the risk of low birth weight and premature births. As births weight is related to education and income outcomes in adulthood (Black et al 2007; Royer 2009; Almond & Currie 2011), the inequalities may translate into the next generation. As such, my results suggest that there are room for additional health insurance for low-income individuals in Denmark. Moreover, as the results are from Denmark, where social insurance is relatively generous and health insurance is universal with relatively low co-insurance, the results could be interpreted as a lower bound for other countries with less generous insurance, as suggested by e.g. Morgan & Lee (2017).

My contribution to the literature is two-fold and relate to both the economic and medical literature. First, a large strand of the economic literature shows how people respond to liquidity constraints and payday. This behavior has been shown for overall consumption as well as across various domains, such as food and caloric intake, fuel, clothing, pharmacies, etc. Furthermore, timing of income and liquidity constraints affect low-income individuals more. See review in Jappelli & Pistaferri (2010) and Fuchs-Schundeln & Hassan (2016).[3] Another strand of the literature shows that payday resets all kinds of activity, such as sport, crime, alcohol and drug consumption, decision-making, and mortality.[4] While the literature has extensively investigated demand across a variety of domains, I am the first to show this within healthcare. Healthcare is a domain, where the marginal benefit of timely consumption may be especially high, as the delaying necessary treatment can have serious health consequences, thus, translating liquidity constraints into direct welfare losses. Further, I show differences across different types of prescription drugs, illustrating the heterogeneity even within this consumption group.

---

[3] Within this literature, related important papers studies overall consumption responses to liquidity. See e.g. Stephens (2003), Shapiro (2005); Stephens (2006); Mastrobuoni & Weinberg (2009); Gelman et al. (2014); Olafsson & Pagel (2018). Another part of the literature focus on evidence from one-time payments, e.g. Parker (1999), Souleles (1999), Shapiro and Slemrod (2003a), Shapiro and Slemrod (2003b), Shapiro and Slemrod (2009), Johnson et al. (2006), Parker et al. (2013), and Broda & Parker (2014); Kreiner et al. (2019).

[4] See e.g. Dobkin & Puller (2007); Foley (2011); Evans & Moore (2011, 2012); Bruich (2014); Andersson et al. (2015); Carvalho et al. (2016); Watson (2019).

Second, an extensive economic and medical literature has shown how patients respond to co-payments for healthcare—especially low-income individuals.[5] I contribute to this literature by showing that liquidity constraints within the month exist. If one is using monthly or yearly data these effects will smooth out, making the effect of liquidity constrains undetectable. Moreover, I show that stockpiling of drugs used to treat chronic conditions is a present mechanism, as previously documented, see e.g. Skipper (2012), Simonsen et al. (2018), and Alpert (2016). Many studies are from the U.S., where both the doctor visit as well as treatment are subject to an out-of-pocket cost for the patient, and cannot as such separate the choice of seeking medical advice from taking the prescribed treatment. Exploiting the Danish health insurance system—with free doctor visits, but co-insurance for prescription drug treatment—I contribute by showing that some individuals do postpone the free doctor visit in the expectation of the subsequent out-of-pocket cost of the prescribed treatment. Furthermore, studies investigating prescription drug consumption in the U.S. normally rely on Medicare Part D, which is only for elderly people. Thus, I contribute with results for prime-age individuals, including pregnant women and parents to young children.

The paper proceeds as follows. In section 2, I describe the institutional setting. In section 3, I describe the data and provide summary statistics. In section 4, I describe the empirical framework. In section 5, I provide the main results and mechanisms. Specifically, section 5.1 analyzes drugs used to treat chronic conditions, while section 5.2 analyzes drugs used to treat acute conditions. In section 6, I discuss and conclude.

## 2. Institutional Background

Denmark has a comprehensive social and health insurance system covering the full population. Welfare benefits is the final safety net for individuals age 18-64, who cannot in any other way support themselves and their family. In 2012, the pre-tax monthly benefit was DKK 10,000 (≈USD 1,600) for a single individual with no children; and DKK 13,000 (≈USD 1,900) with children. The plan is income- and means-tested, so the individual cannot have any other income nor wealth exceeding DKK 10,000 (≈USD 1,600), including the value of house car, etc.[6] The benefits are postpaid and paid out monthly—directly into the recipients bank account—on the last business day of the month.[7]

---

[5] See e.g. DiMatteo (2004); Krueger et al. (2005); Osterberg & Blaschke (2005); Chandra et al. (2010); Finkelstein et al. (2012); Baicker et al. (2013); Einav et al. (2015); Baicker et al. (2013); Brot-Goldberg et al. (2017); Abaluck et al. (2018); Watson et al (2019).
[6] See Appendix Table 3 for additional information on the transfer payments.

Health insurance in Denmark is a universal, tax-financed, single-payer system run by the government. Most services are free-of-charge for the individual (e.g. physician consultations and hospital visits), while other services are subsidized generously (e.g. physiotherapy, dentistry, and prescription drugs). For prescription drugs, Denmark has a progressive piece-wise linear subsidy system for the yearly accumulated expenditures, so that patients with higher expenditures receive more subsidy.[8] The subsidy is subtracted from the total price at the time of purchase at the pharmacy, so the patient only have to pay her out-of-pocket share. In 2012, the co-insurance rates were as follows: 100% co-insurance for the first DKK 890 (≈USD 140), 50% until DKK 1,450 (≈USD 220), 25% until DKK 3,130 (≈USD 480), and 15% above.[9]

## 3. Data and Summary Statistics

I combine several administrative data sources, linked via person-level identifiers. All residents in Denmark have a unique personal identification number that allows a completely accurate linkage of information across different registers at the individual level. I use the Government Transfer Registers to construct my main population by identifying all individuals receiving welfare benefits from 2000 to 2019. The *Prescription Drug Registers* comprise all prescription drug purchases, including detailed information on the product, prices and date of purchase. The Anatomical Therapeutic Chemical (ATC) Classification System allows me to identify different types of drugs, e.g. antibiotics and cholesterol-lowering statins. I link family members via the *Family Registers*. This allows me to identify prescription drug consumption children of mothers receiving welfare benefits. Lastly, for a subsample (2014 - 2016), I can identify the date of the physician writing the prescription. Together with the day of the purchase, I construct a proxy for how many days patients postpone filling their prescription.

The time period for my sample is March 18, 2000, to November 14, 2016. This leaves me with 200 payday events. For each payday event, I construct a 28-day event period—13 days before and 15 days after. I restrict my sample to only include individuals that receive cash welfare for the full 28-day month, in order to have a balanced sample per payday event. Table 1 provides summary statistics on the main sample. The sample consists of 768,625 unique individuals over the 17-year period with a total of 2,052,113 spells—averaging 100,000 individuals per payday event. 44% are women, the mean age is 31, and the mean number of spells is 3.9. The average length of the first spells is 59 weeks, with men having shorter spells (48 weeks compared to 62 for women).

---

[7] This leads to 43% of paydays falling on Fridays. Appendix Table 2 summarizes the distribution of paydays on weekdays.
[8] Appendix Figure 10 illustrates the co-insurance scheme.
[9] For more details on the institutional background of the Danish co-insurance scheme for prescription drugs, see e.g. Simonsen et al. (2016).

*Table 1 here*

On the limitations of the data, note, I only observe purchases of prescription drugs. This means that I do not observe prescription that are never filled, drugs administered directly at the hospital, over-the-counter drugs, nor if the patient actually consumes the drugs. This additional challenge to adherence is beyond the scope of this paper.[10]

## 4. Empirical Framework

*Payday Sensitivity*

To analyze the payday sensitivity of filling prescriptions, I employ an event study approach. Specifically, I estimate the following regression:

$$y_{i,t} = \sum_{\tau=-12}^{14} I_\tau \times \beta_\tau + \alpha_i + \delta_{Weekday} + \psi_{specialday} + \gamma_{year} + \mu_{month} + \varepsilon_{i,t}, \quad (1)$$

where $y_{i,t} = 1$ if individual $i$ fills a prescription on date $t$ and zero otherwise; $\tau$ is day relative to transfer income payday ($\tau = 0$), which runs from 13 days before to 14 days after payday, excluding $\tau = -13$; $I_\tau$ are indicators of days relative to payday. $\alpha_i$ are individual fixed effects, $\delta_{Weekday}$ are day-of-the-week fixed effects (Monday through Sunday), $\psi_{specialday}$ are fixed effects for special days (e.g. New Years Eve, May First, and bank holidays),[11] $\gamma_{year}$ are year fixed effects, and $\mu_{month}$ are payday event month fixed effects.[12]

The 27 $\hat{\beta}_\tau$ estimates are the parameters of interest, measuring the *absolute* difference in propensity to fill at least one prescription on day $\tau$ relative to $\tau = -13$. This can be interpreted as the change in the probability that an individual will fill a prescription on day $\tau$ relative to an average day in the middle of the month—in percentage points.

---

[10] For the interested reader, Pottegård et al. (2014) provides some evidence of primary non-adherence.
[11] Appendix Table 3 provides a complete list of special days used as controls.
[12] Note, I do not have to leave out two days in the regression, when including individual fixed effects, as I have many events per individual and the distance between the events varies across individuals, because individuals have variation in the number and length of spells.

I normalize the standard event study statistic $\hat{\beta}_\tau$ with the average of the dependent variable (i.e., the baseline), in order to ease comparison across different drug and population groups with different baselines:

$$\tilde{\beta}_\tau = \hat{\beta}_\tau / \bar{y}, \qquad (2)$$

where $\bar{y}$ is the average filling rate of the drug under investigation.

The 27 $\tilde{\beta}_\tau$ estimates measure the *relative* change in propensity to fill a prescription on day $\tau$ relative to an average—in percent.

*Test for liquidity constraints*

To test for liquidity constraints being the driving mechanism, I estimate the probability of postponing filling the prescription exactly *x* days by estimating equation:

$$y_{i,t} = \sum_{\tau=-12, s\neq 7}^{14} I_\tau \times \beta_\tau + \alpha X_{i,t} + \delta_{DOW} + \psi_{specialday} + \gamma_{year} + \mu_{month} + \varepsilon_{i,t}, \qquad (3)$$

where $y_{i,t} = 1$ if individual *i* postpone filling the prescription *x* days and zero otherwise, conditional on getting the prescription on day *t*. I estimate the equation 14 times, i.e., where $y_{i,t}$ is a dummy for postponing zero days, one day, two days, etc. $X_{i,t}$ controls for age, sex and age-by-sex.

## 5. Main Results and Mechanisms

In this section, I provide the main results and heterogeneity analyses to inform about the mechanisms driving the results. First, I provide evidence of the overall payday sensitivity for filling *any type* of drug. Next, I study mechanisms driving this behavior. In this analysis, I exploit that some types of drugs are used repeatedly over a longer time, e.g. prophylaxis drugs for chronic conditions. For these types of drugs, a large payday sensitivity is consistent with stockpiling behavior being a driving mechanism. Another potential important mechanism is liquidity constraints. To investigate if this mechanism is a key operating mechanism in this setting, I study acute antibiotic treatment, where the patient cannot anticipate the timing of disease. To support this mechanism of liquidity constraints, I exploit a unique feature of the data allowing me to identify the exact difference in timing between the physician writing the antibiotic prescription and the patient filling it. To study the severity of these behavior, I study heterogeneity of vulnerable groups—pregnant women and children—who have potential higher health consequences of postponing antibiotic treatment.

**5.1 Any type of drug**

First, I group all types of drugs together. On average, an individual on welfare benefits has a propensity of 1.3% to fill a prescription on any given day. Figure 1 illustrates the response to transfer income payday ($\tau = 0$) relative to 13 days before payday. On payday, there is a 52% increase in the propensity to fill a prescription; the day after, there is an increase of 33%; and two days after, there is an increase of 11%. The increase fades out over the following days. Moreover, there is a clear decrease of 4-8% in the four days preceding payday. However, these overall effects include a bundle of different types of drugs, e.g. stockable drugs (where the patient can anticipate the need to fill the prescription) as well as drugs for unanticipated treatment. Thus, drugs used to treat chronic conditions, where the patients are able to stockpile and still smooth actual consumption, can potentially drive the overall response illustrated in Figure in 1.[13]

*Figure 1 here*

**5.2. Drugs for Chronic Conditions**

To investigate a potential mechanism of stockpiling behavior, which has been documented by several studies (Skipper 2012, Gelman et al. 2014, Simonsen et al. 2018, Einav et al. 2015, Alpert 2016, Brot-Goldberg et al. 2017), I study drugs used to treat chronic conditions. I separate drugs, where patients should be able to anticipate the need to fill the prescription. I focus on six highly predictable drug groups. That is birth control pills and medication for five chronic diseases: ace inhibitors for high blood pressure, beta blockers for cardiac arrhythmia, oral anti-diabetes drugs for type 2 diabetes, statins for high cholesterol, and antipsychotics for Schizophrenia. All six types of drugs are taken for a longer period (most for the rest of your life) and are possible to stockpile, why the need to fill the prescriptions should be anticipated by the patient. Figure 2 presents the payday sensitivity for each of the six types of drugs. With an increase of 52-99% at payday, the payday sensitivity is clearly larger for these types of drugs, which is also, what one ex-ante would have expected. Again, I find a similar pattern of decrease in the four days preceding payday.

*Figure 2 here*

**5.3. Drugs for Acute Conditions**

To investigate a potential mechanism of liquidity constraints, I study drugs used to treat acute conditions. Studying drugs used to treat conditions acute conditions that are uncorrelated with payday and where treatment should not be postponed can reveal if individuals are liquidity constrained. I focus on antibiotics, which are used to treat microbial infections. When treated promptly and properly, infections rarely lead to complications. Prevalence, symptoms and complications vary across different infections. For instance, the most common disease treated by antibiotics is urinary tract infections (UTI). If left untreated, it can have serious health consequences, such as recurrent infections can lead to permanent kidney damage, increased risk in pregnant women of delivering low birth weight children or having premature birth, urethral narrowing in men, or sepsis—a potentially life-threatening complication of an infection (Mayo Clinic 2019b). For the frequent disease of pneumonia, the risk of complications increase, if treatment is not started promptly—especially for patients in high-risk groups: infants and young children, elderly people, and people with health problems or weakened immune systems. One complication is bacteria in the bloodstream (bacteremia), i.e., bacteria that enter the bloodstream from the lungs can spread the infection to other organs, potentially causing organ failure (Mayo Clinic 2019a).[14]

---

[13] Appendix Figure 4 provides a robustness check on the payday weekday, e.g. by excluding all payday event months, where the payday falls on a Friday.

Figure 3 presents the payday sensitivity for filling antibiotic prescriptions. Even for these types of drugs—where there should be no anticipation nor stockpiling—I find a similar behavior of payday sensitivity. On payday, there is an increase of 22% in the propensity to fill an antibiotic prescription; and in the four days preceding payday, there is a decrease of 5-8%. This suggests that individuals on welfare benefits postpone treatment that should not be postponed.

<div align="center">*Figure 3 here*</div>

### 5.3.1. Postponement of treatment

My data provides the unique opportunity to investigate how severe the problem of postponing antibiotic treatment is. I leverage a subsample of the data (2014-2016), where data on visits to the general practitioner (GP) is recorded on a daily basis.[15] This information allows me to track all antibiotic prescriptions back in time to the last GP visit, linking the prescription on patient and provider ID.[16] 86% of all antibiotic prescriptions are prescribed by a GP and 10% by a hospital physician.[17] GPs do not have a stock of prescription drugs nor a pharmacy in the clinic; hence, the patient have to go to a pharmacy to fill the prescription.

---

[14] Appendix table 6 provides summary statistics for the indication ('diagnosis') for the antibiotics prescriptions. 25% of all antibiotic prescriptions are against urinary tract infections (UTI);[14] 9.7% are against skin and soft tissue infections, i.e., an open wound that caught infection; 9.7% are against pneumonia; 1.2% are against Chlamydia trachomatis, which is the most common sexually transmitted infection (STI); and 14% of the prescriptions have no indication filled out.
[15] Prior to June 23, 2014, GP visits are only recorded on the weekly basis.
[16] With this approach, I am able to link 70% of all antibiotic prescriptions prescribed by a GP, amounting to 9.3 million antibiotic prescriptions to 3.1 million unique patients over the five-year period.
[17] The remaining 4% are either invalid or prescribed from outside Denmark.

Postponement of doctor visits

A natural question ask is whether this response is driven by patients postponing filling the prescription or by patients postponing the visit to the doctor in the expectation of the subsequent out-of-pocket cost for the prescribed treatment. To investigate this latter hypothesis, I estimate equation (1), where I use the date of the doctor visit relative to payday (instead of the date of filling the prescription). Figure 4 presents the results. On payday, there is a 10% increase in the number of doctor visits. This suggests that some patients on welfare benefits do postpone the (free) doctor visit due to the expectation of subsequent out-of-pocket costs related to the prescribed treatment. However, it cannot account for the entire timing effect of antibiotic prescription fillings presented in Figure 4, which was 22%.

*Figure 4 here*

Postponement of filling antibiotic prescriptions

Next, I exploiting the difference in timing between the doctor writing the prescription and the patient filling it to estimate the number of days each patient postpone filling their antibiotic prescription.[18] Figure 3 shows that 80% cash welfare recipients fill their antibiotic prescription within the same day as the doctor writing it (t=0), 90% within the day after (t=1); and 98% fill it within 10 days.

*Figure 5 here*

Test for Liquidity Constraints as the Driving Mechanism

To test for liquidity constraints being a key operating mechanism, I combine the information on the day of the doctor writing the prescription with the number of days the patient postpone filling the prescription. If monthly liquidity constraints is an operating mechanism, one would expect that patients getting the prescription *one* day before payday would postpone filling the prescription *one* day, patients getting a prescription *two* days before payday would postpone filling the prescription *two* days, etc. Moreover, one would expect the response to be stronger in the days just before payday, as liquidity constraints are more binding by the end of the month and the health consequences increase with the number of days delaying treatment.

To test this hypothesis, I construct a dummy for the exact number of days postponing filling the prescription and estimate equation (3), where the distance to payday is the day the doctor writes the prescription. For instance, let $y_{i,t} = 1$ if the patient postpone filling the prescription *one* day and let $y_{i,t} = 0$ otherwise.

---

[18] This approach relies crucially on patients not having any other contact with the GP between the date of the physician writing the prescription and the date of filling the prescription.

Figure 6, panel (a) shows the propensity to postpone filling the prescription *zero* days, i.e., filling the prescription the same day as the doctor writing it. There is no sensitivity on payday or the following days. However, there is a significant decrease in the days preceding payday—with a decrease of 10 pp on the day preceding payday. Panel (b) shows the propensity to postpone filling the prescription *one* day. Individuals getting the prescription *one* day before payday has an increase of 7 pp in the propensity to postpone treatment exactly *one* day. Panel (c) shows the propensity to postpone *two* days. Individuals getting the prescription *two* days before payday has an increase of 5 pp in the propensity to postpone treatment exactly *two* days. Panel (d)-(n) shows the behavior for postponing three to 11 days. The response decreases with the distance to payday and fades out when there is more than eight days to payday.

These results are consistent with liquidity constraints being a key operating mechanism for individuals on welfare postponing antibiotic treatment, as patients do seek medical attention by seeing the doctor (which is free-of-charge), but thereafter postpone the prescribed treatment (that comes with an out-of-pocket price) until transfer income payday.

Another well-known mechanism that have been posed by the payday literature is that payday may be a day in which people get out and take care of a variety of household business and social tasks, e.g. due to fixed costs of going out on an excursion. However, as the individuals do see a doctor, when they get a prescription, they have already paid the fixed cost of leaving the house. Hence, the choice of not filling the prescription right away is unlikely to be due to this proposed mechanism.

*Figure 6 here*

**Heterogeneity in Health Cost**

Economic theory predicts that individuals with a higher (health) cost of postponing treatment should have a smaller payday sensitivity. In the case of antibiotic treatment, pregnant women and children are vulnerable groups (Longo et al. 2013; Mayo Clinic 2019a).[19]

Figure 8, Panel (a) shows that even for pregnant women, I find an increase in the propensity to fill an antibiotic prescription of 19% and a decrease in the four days preceding payday. Panel (b)

---
[19] Other vulnerable groups include elderly and patients with health problems (e.g. diabetes and alcoholism) or weakened immune systems (e.g. cancer and lupus patients).

presents no clear evidence for children below age 13, where the mother is on welfare, suggesting that the payday sensitivity does not spillover to the children.

Interestingly, I do find a payday sensitivity behavior for both pregnant women and children, when looking at *any type* of drug, as presented in Appendix Figure 16, indicating that stockpiling may still be present for other types of drugs prescribed to children.

*Figure 7 here*

## 6. Discussion and Conclusion

I show that welfare recipients living off social insurance are liquidity constrained by the end of the month to a degree that they postpone filling necessary prescriptions. I find an overall effect for *any type* of drug of a 52% increase on payday. To disentangle potential mechanisms, I separate different types of drugs. First, I separate prophylaxis drugs used to treat chronic conditions, where the patient has to fill the prescription recurrently. For these types of drugs, I find a 52-99% increase in the propensity to fill a prescription on payday, implying that stockpiling is an operating mechanism.

Second, I separate drugs used to treat acute conditions, where treatment should not be delayed. Even for these types of drugs, I find a 22% increase at payday. I show that if individuals get the prescription *one* day before payday, they postpone filling the prescription *one* day; and, that if individuals get the prescription *two* days before payday, they postpone filling the prescription *two* days. This documents that monthly liquidity constraints is a key operating mechanism for why individuals living paycheck-to-paycheck postpone starting necessary treatment, when they are faced with out-of-pocket expenditures.

Lastly, I separate vulnerable patient groups, for whom the health cost of postponing acute treatment is especially high. Even for pregnant women, I find a behavior of a payday effect with an increase of 19%. That the pregnant women postpone antibiotic treatment can in itself have health consequences for the mother, but may also affect the fetus by increasing the risk of low birth weight and premature birth and thereby translate into non-health outcomes for the next generation. An extensive literature has documented that conditions during in utero can affect long-term outcomes, i.e., the fetal origins hypothesis (for an overview of the empirical works, see Almond & Currie 2011; Almond et al. 2018). For instance, Black et al (2007) find that a 10% increase in the birth weight increase high school completion by 0.9 pp (1.2%), while Royer (2009) finds that a 10% increase (250-gram) increase in the birth weight increases schooling by 0.04. This suggests that if a pregnant woman living of welfare benefits catches a bacterial infection in the days preceding payday, she might postpone treatment until transfer income payday. This leads to the infection being active and potentially growing for a longer time, which can affect the fetus. Hence, liquidity constraints for pregnant mothers may translate into health inequalities already in utero, affecting the next generation.

My results suggest that there are room for additional health insurance for low-income individuals in Denmark. Moreover, as the results are from Denmark, where social insurance is relatively generous and health insurance is universal with relatively low co-insurance, the results could be interpreted as a lower bound for other countries with less generous insurance, as suggested by e.g. Morgan & Lee (2017).

# Figures

**Figure 1: Propensity to fill a prescription for any type of drug**

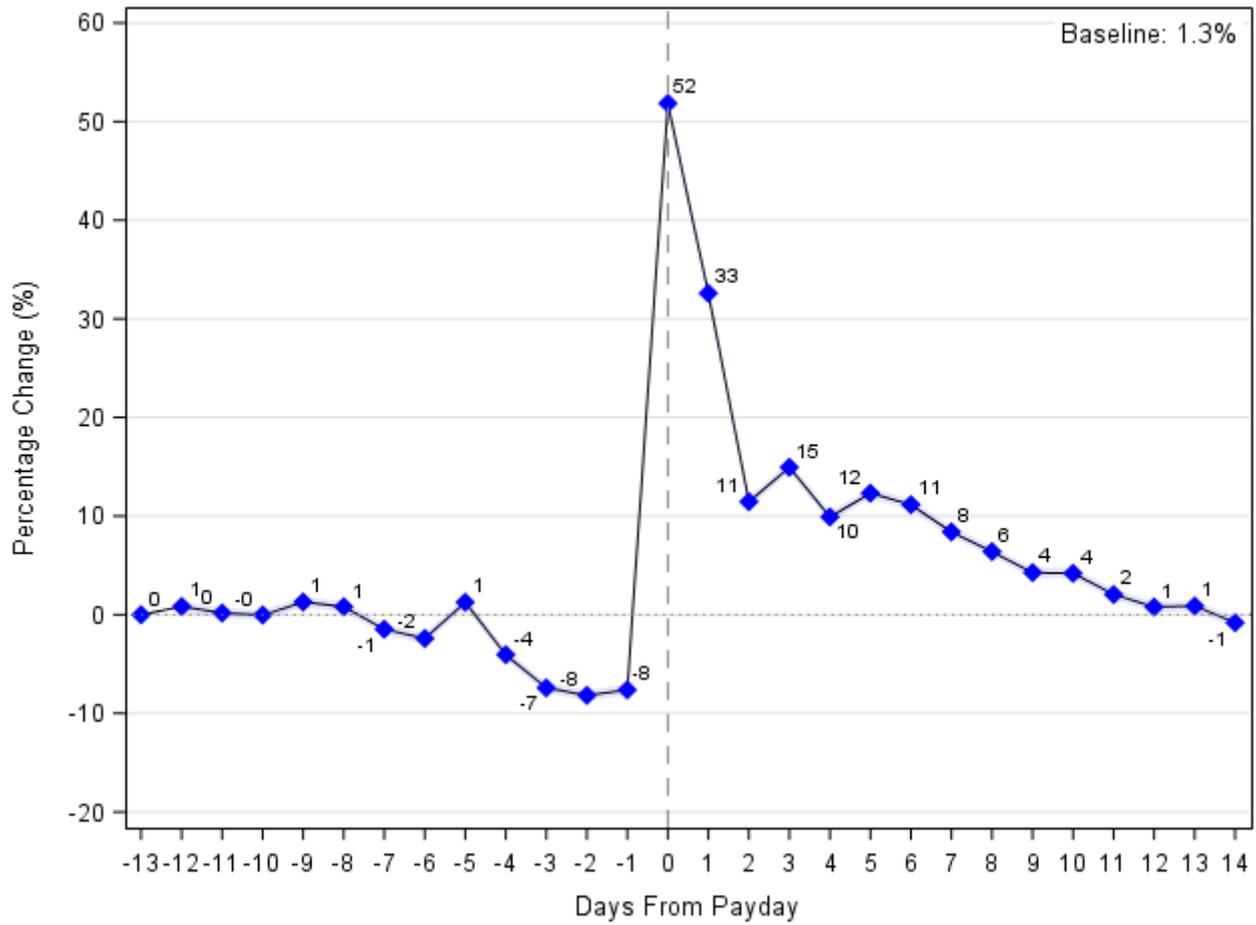

Notes: The figure illustrates the change in propensity to purchase a prescription drug on each day relative to payday (t=0), on a baseline of 1.3%. Welfare recipients have a 52% increase in the propensity to fill a prescription on transfer income payday (t=0). The figure illustrates the $\tilde{\beta}_\tau$ estimates from equation (1) and normalized by the mean as in equation (2). Shaded areas are 95% confidence bands.

# Figure 2: Propensity to Fill a Prescription for a Chronic Condition

(a) Ace Inhibitors (high blood pressure)

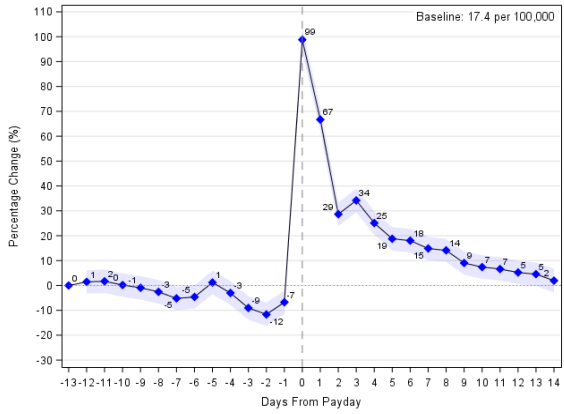

(b) Beta Blockers (cardiac arrhythmia)

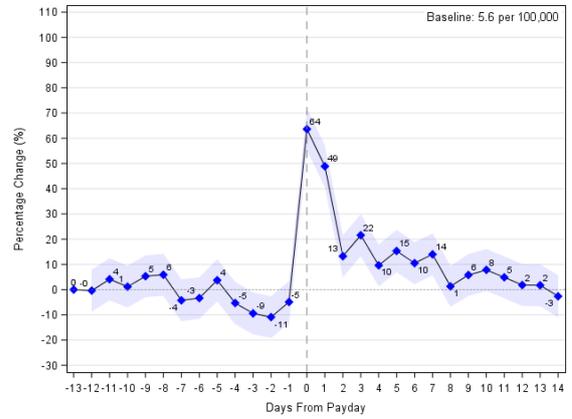

(c) Oral Anti-diabetes Drugs (type-2 diabetes)

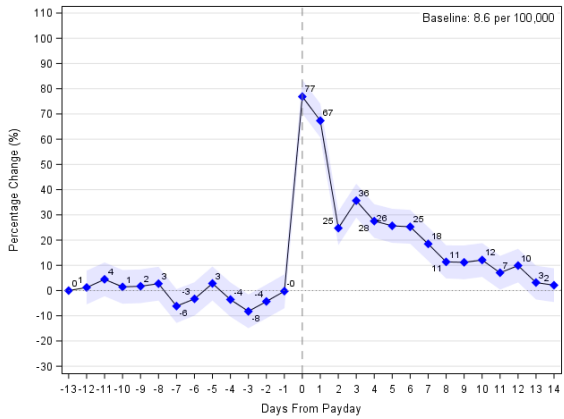

(d) Statins (high cholesterol)

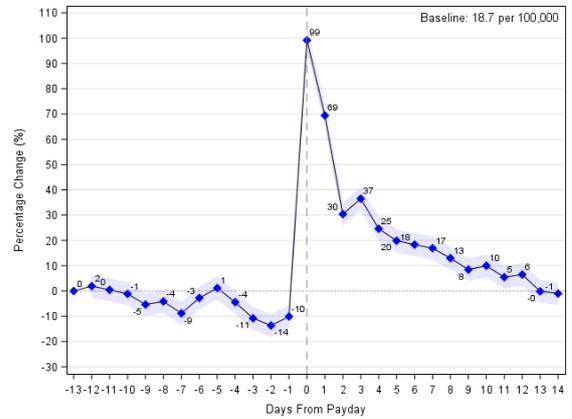

(e) Antipsychotics (Schizophrenia)

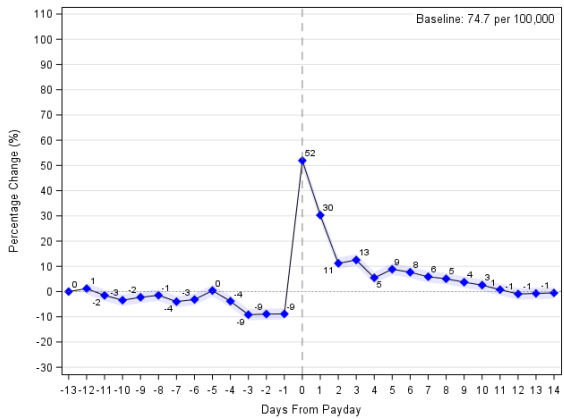

(f) Birth Control Pills

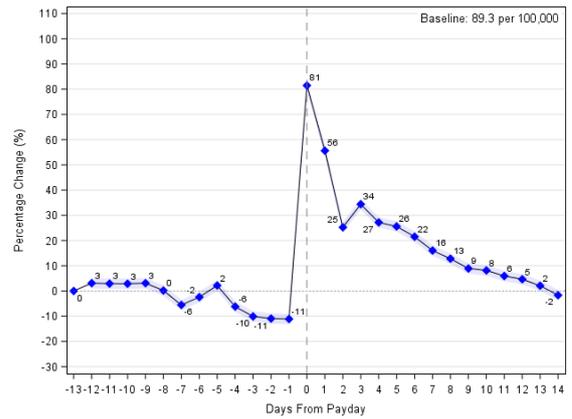

Notes: The figure illustrates the change in the propensity to purchase six common drug types used to treat chronic conditions on each day relative to payday (t=0). The figures illustrates the $\tilde{\beta}_\tau$ estimates from equation (1) and normalized by the mean as in equation (2). Shaded areas are 95% confidence bands. See Appendix Table 4 for the ATC codes used to identify each drug group.

# Figure 3: Propensity to Fill an Antibiotic Prescription

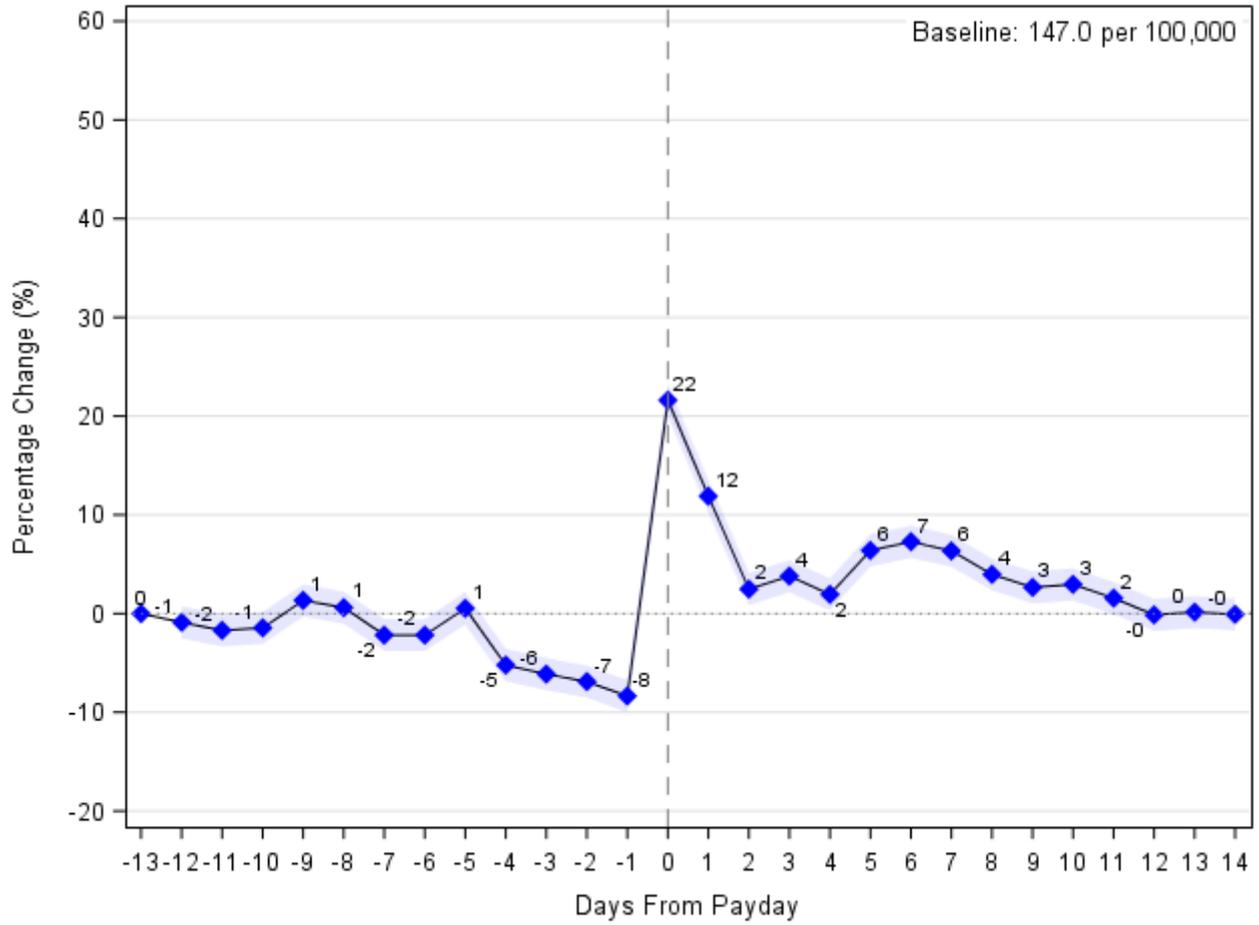

Notes: The figure illustrates the change in the propensity to purchase antibiotics on each day relative to payday (t=0). The figure illustrates the $\tilde{\beta}_\tau$ estimates from equation (1) and normalized by the mean as in equation (2). See Appendix Table 5 for the ATC codes used to identify the drug group. Shaded areas are 95% confidence bands.

# Figure 4: Doctor Visits Relative to Payday

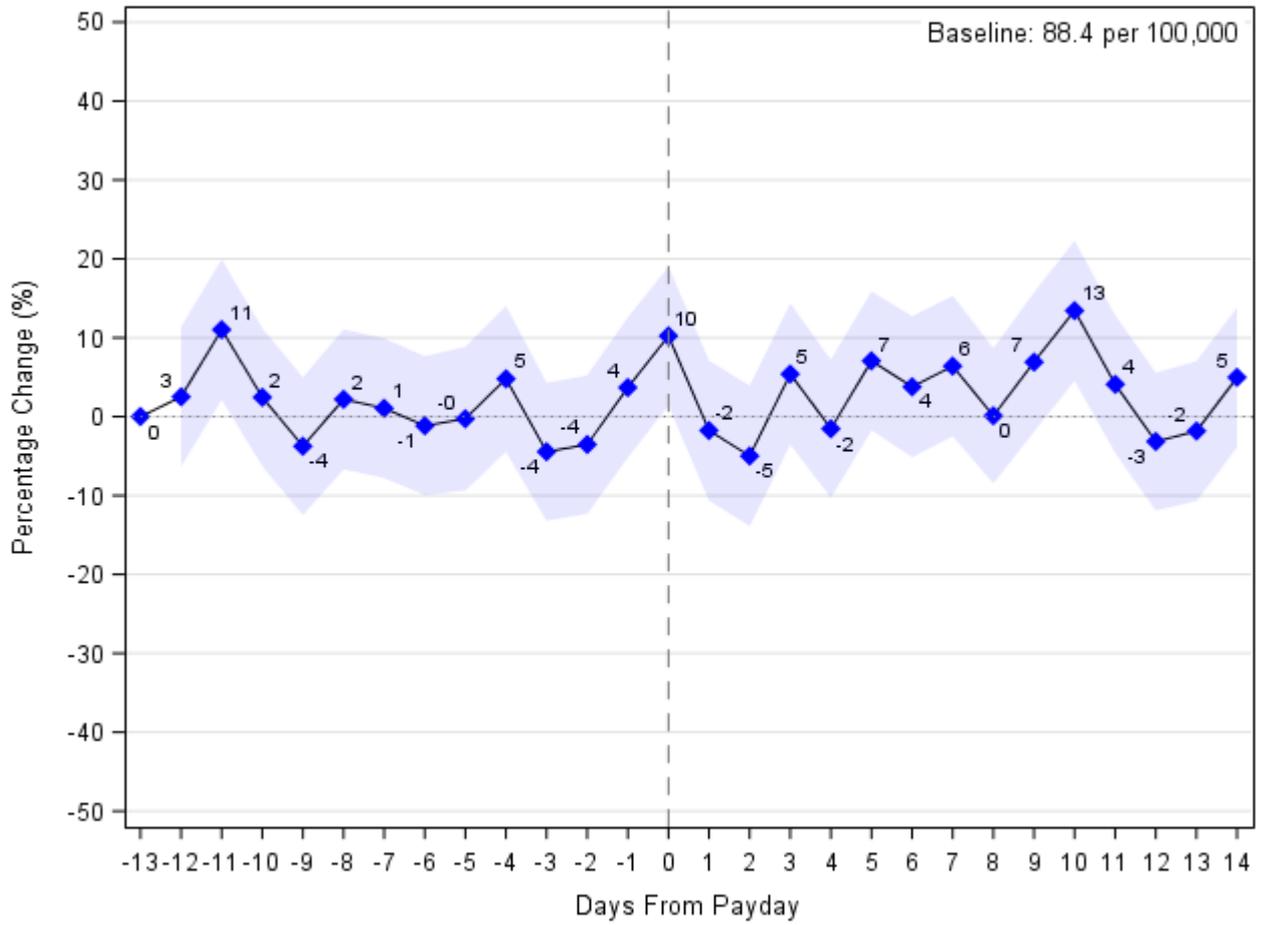

Notes: The figure illustrates the change in the propensity to go the doctor to get an antibiotic prescription relative to payday (t=0). The figures illustrates the $\tilde{\beta}_\tau$ estimates from equation (1) and normalized by the mean as in equation (2). Shaded areas are 95% confidence bands.

**Figure 5: Number of Days Postponing Filling Antibiotic Prescriptions**

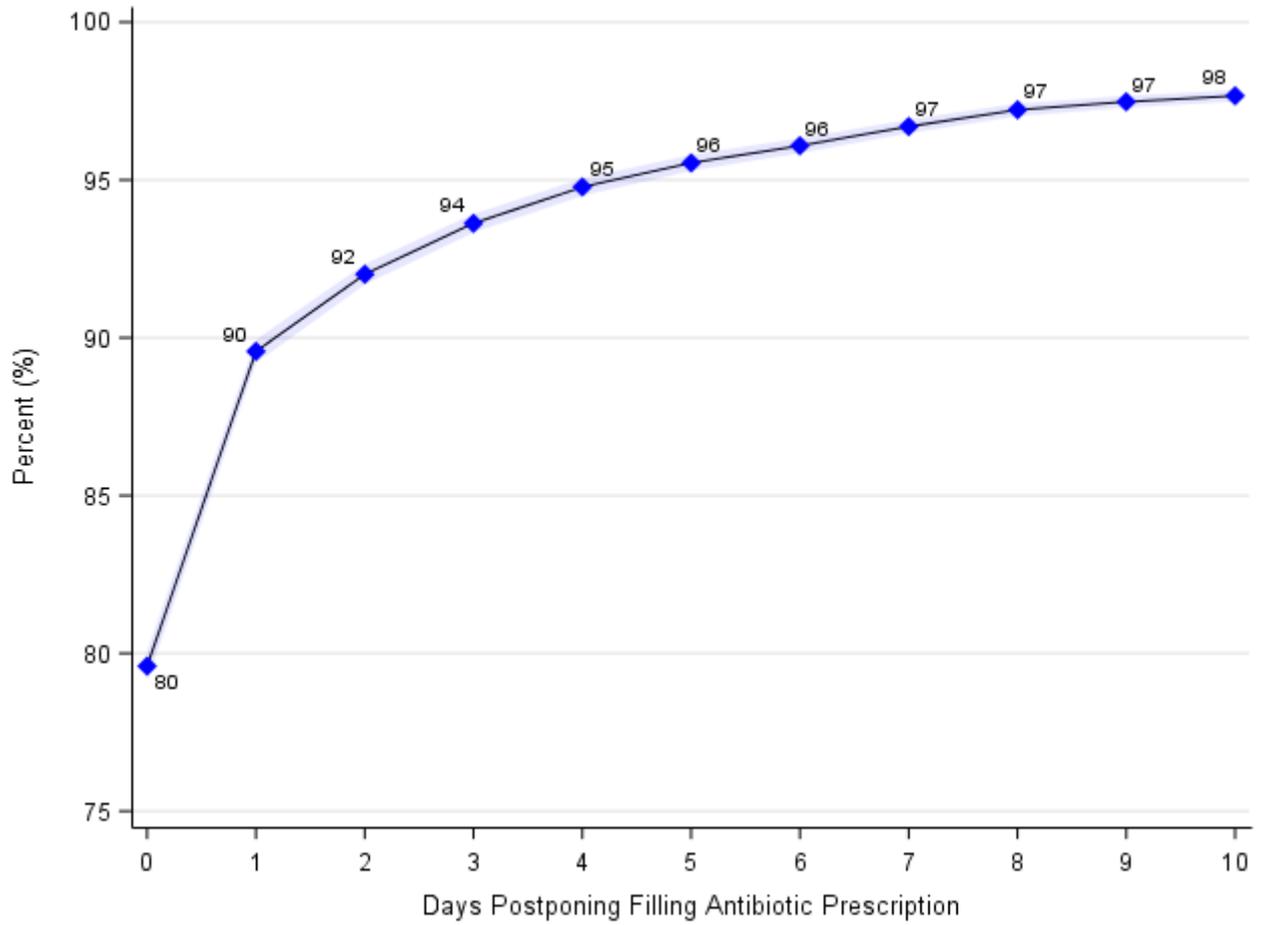

Notes: The figure illustrates how many days individuals on welfare benefits postpone filling their antibiotic prescription (conditional on filling it within 30 days). 80% fill it within the same day as the doctor writing it (t=0), 90% fill it within the day after (t=1), and 98% fill it within 10 days. Shaded areas are 95% confidence bands.

# Figure 6: Days Postponing Filling the Antibiotic Prescription

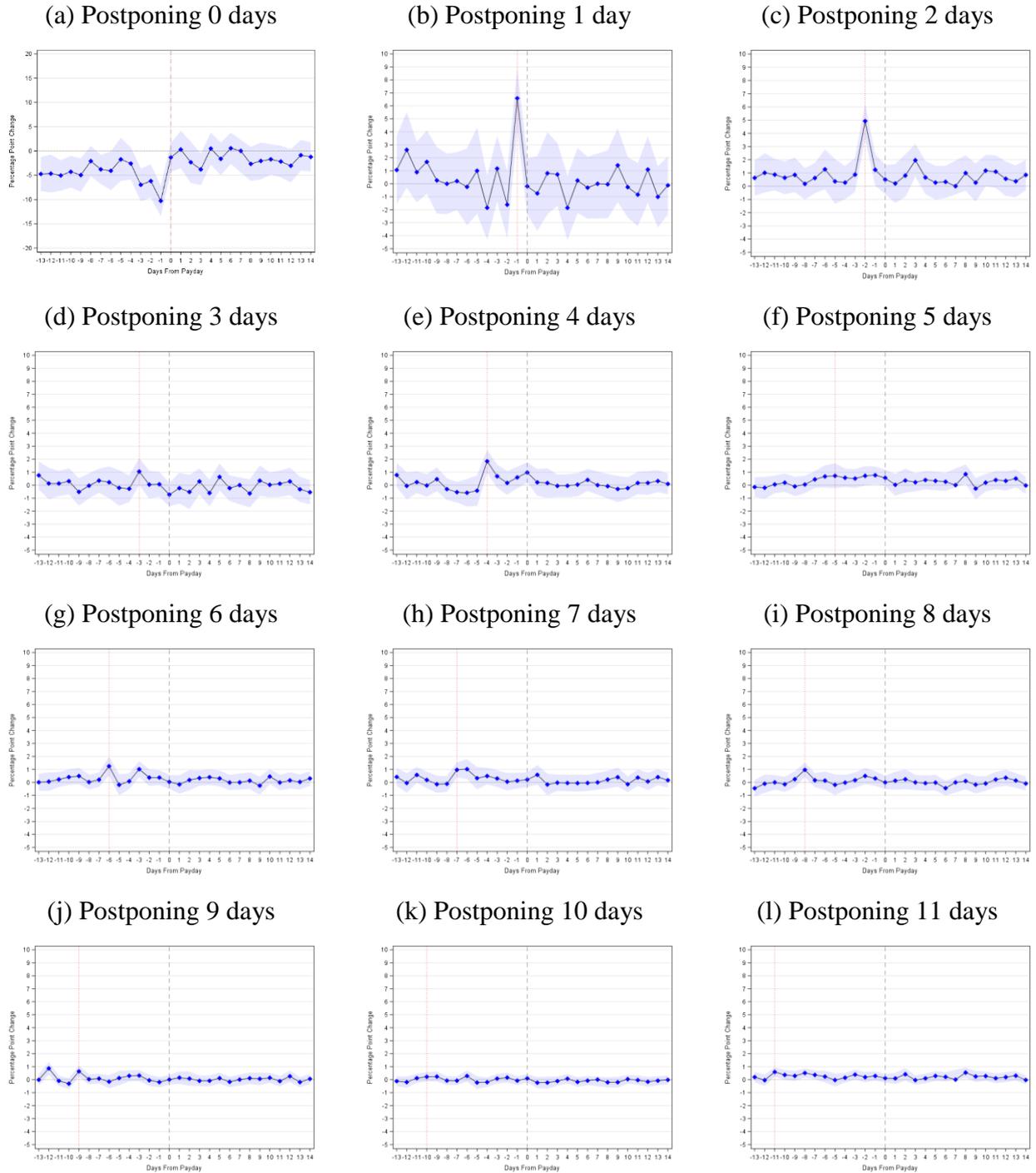

Notes: The figure illustrates the change in propensity to postpone filling the prescription 0-11 days. Panel (a) shows the propensity to postpone filling the prescription zero days, i.e., fill the prescription the same day as the doctor writing it. Panel (b) shows the propensity to postpone filling the prescription one day. For individuals getting the prescription one day before payday (t=-1), there is an increased propensity to fill the prescription one day after. Panel (c) shows the propensity to postpone filling the prescription two days. For individuals getting the prescription two days before payday (t=-2), there is an increased propensity to fill the prescription after two days. Shaded areas are 95% confidence bands.

# Figure 7: Propensity to Fill an Antibiotic Prescription for Vulnerable Patients

(a) Pregnant Women  (b) Children

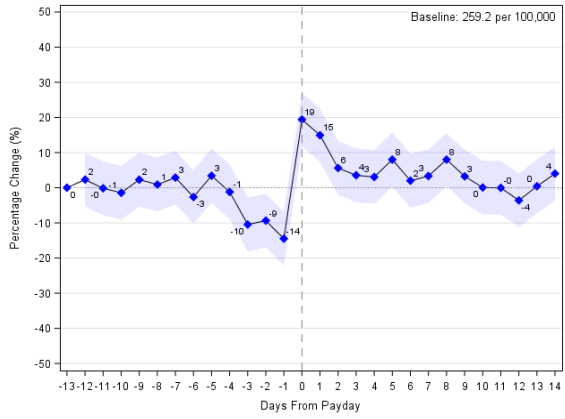 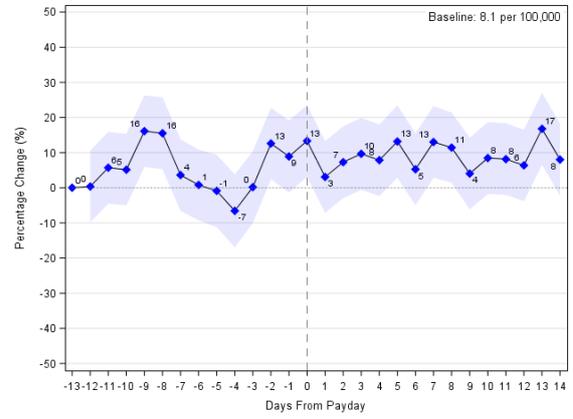

Notes: The figure illustrates the change in the propensity to purchase antibiotics on each day relative to payday (t=0). The figure illustrates the $\tilde{\beta}_\tau$ estimates from equation (1) and normalized by the mean as in equation (2). See Appendix Table 5 for the ATC codes used to identify the drug group. Panel (a) illustrates the response for pregnant women on welfare benefits and Panel (b) the response for children younger than age 13, whose mother is on welfare benefits. Shaded areas are 95% confidence bands.

# Tables

## Table 1: Summary Statistics

|  | Overall | Females | Males |
|---|---|---|---|
| Number of individuals | 768,625 | 370,171 | 398,454 |
| Female share (%) | 43.8 | - | - |
| Age, mean | 31.2 | 31.1 | 31.6 |
| **Spells** | | | |
| ***Number of spells*** | 2,052,113 | 899,686 | 1,152,427 |
| Median | 2 | 2 | 3 |
| Mean | 3.9 | 3.6 | 4.1 |
| ***Length of spells*** *(weeks)* | | | |
| <u>First spell</u> | | | |
| Median | 19 | 20 | 18 |
| Mean | 59.0 | 66.6 | 52.1 |
| Standard Deviation | 105.1 | 118.5 | 89.9 |
| <u>All spells</u> | | | |
| Median | 18 | 20 | 17 |
| Mean | 53.7 | 61.6 | 47.5 |
| Standard Deviation | 94.0 | 106.7 | 82.3 |

Notes: This table provides summary statistics on the main sample population. The spells refers to uninterrupted spells on welfare benefits.

# Appendix Figures

**Appendix Figure 1: Co-insurance Scheme for Prescription Drugs**

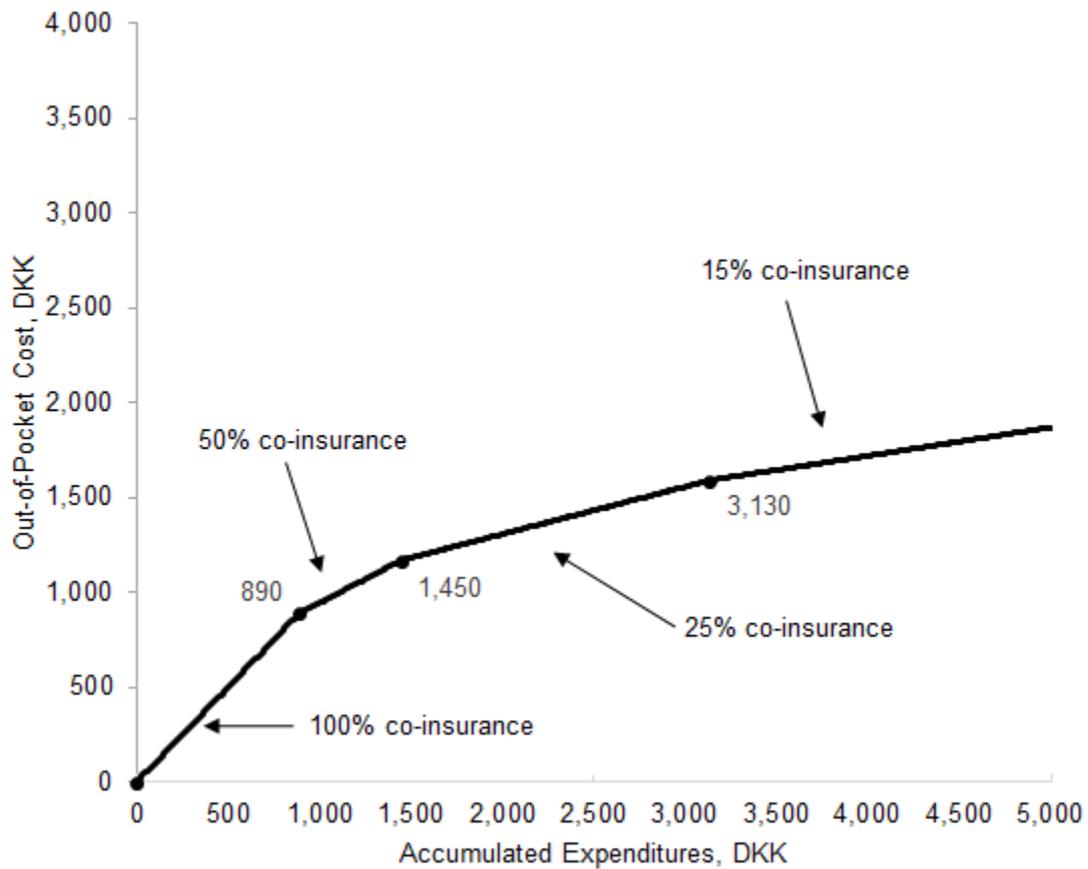

Notes: This figure illustrates the universal co-insurance scheme for prescription drugs in Denmark, 2012. There are 100% co-insurance for the first DKK 890 (i.e., the patient pays the full cost out-of-pocket), 50% co-insurance for the next DKK 560, 25% co-insurance for the next DKK 1,680, and 15% co-insurance for the rest of the subsidy year. The accumulated drug consumption is calculated as a running year, starting at the day of the first purchase and running until the same day one year later (e.g. if one purchase a prescription on May 15, 2002, she resets her balance on May 15, 2003.

**Appendix Figure 3: Propensity to Fill an Antibiotic Prescription**

(a) Narrow-spectrum                    (b) Broad-spectrum

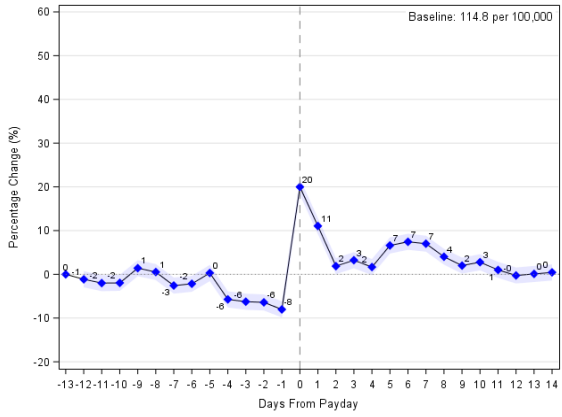 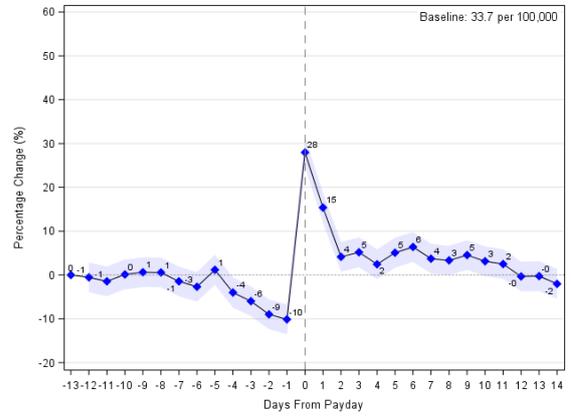

Notes: This figure illustrates the change in the propensity to fill an antibiotic prescription divided on narrow- and broad-spectrum, i.e., splitting the behavior illustrated in Figure 3. See Appendix Table 5 for the ATC codes used to identify the drug groups. Shaded areas are 95% confidence bands.

**Appendix Figure 4: Robustness for Weekday of Payday**

(a) No Exclusion

(b) Excluding Payday on Mondays

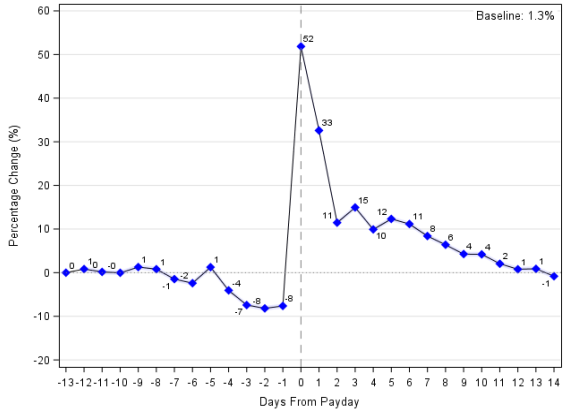
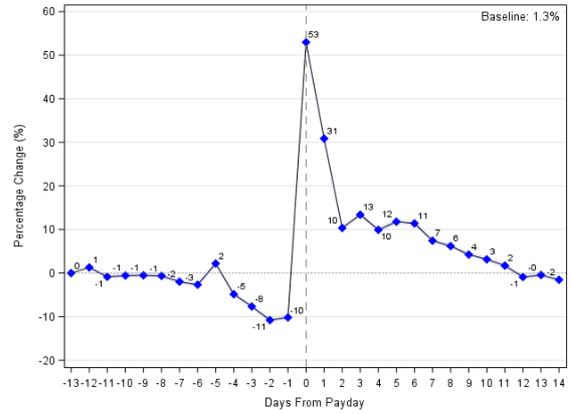

(c) Excluding Payday on Tuesdays

(d) Excluding Payday on Wednesdays

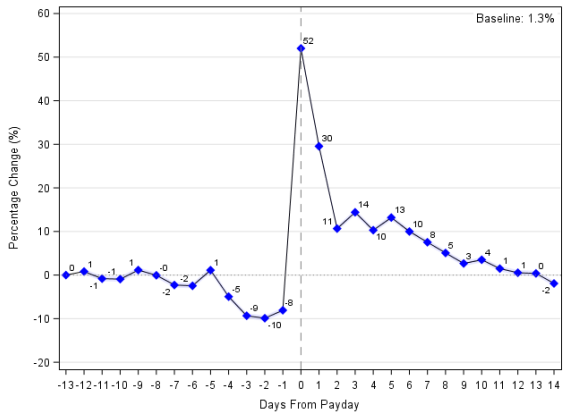
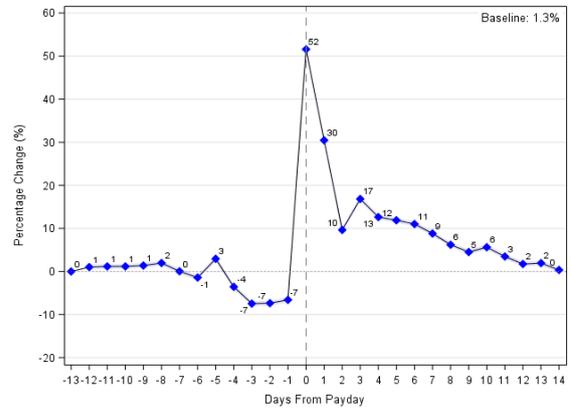

(e) Excluding Payday on Thursdays

(f) Excluding Payday on Fridays

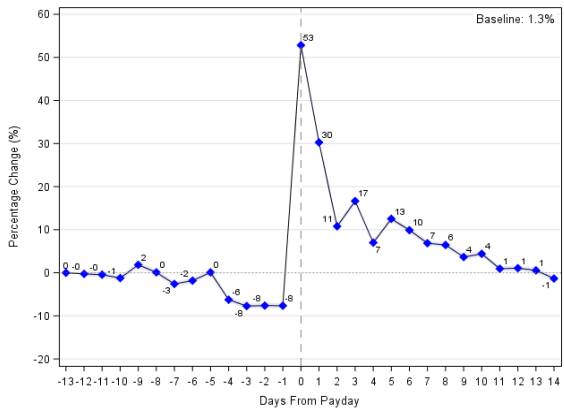
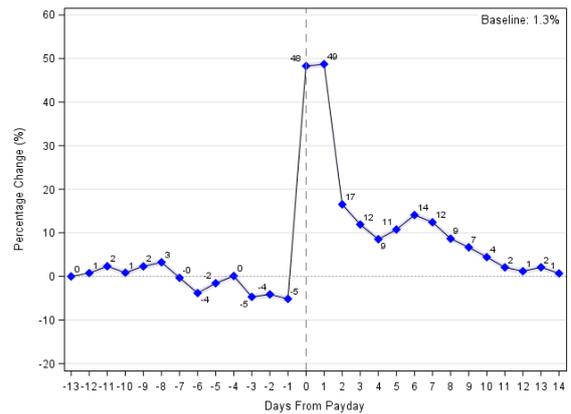

Notes: This figure provides evidence on the sensitivity of the day of the week that payday falls on. Panel (a) provides the pooled estimate, i.e. the same as Figure 1. Panel (b) exclude payday event months, where payday is falling on a Monday. Panel (f) exclude payday event months, where payday is falling on a Monday, which is 42% of all payday events. See Appendix Table 2 for the weekday distribution of paydays. Shaded areas are 95% confidence bands.

**Appendix Figure 5: Propensity to fill a prescription for any type of drug for Vulnerable Patients**

(a) Pregnant Women  (b) Children

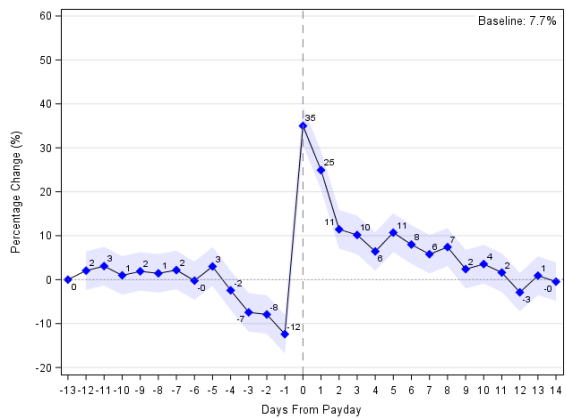
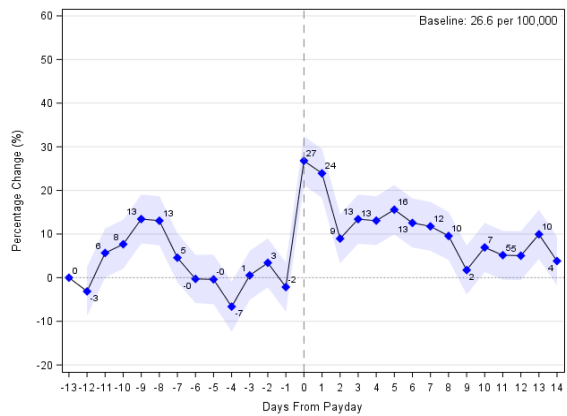

Notes: The figure illustrates the change in propensity to purchase a prescription drug on each day relative to payday (t=0). The figure illustrates the $\tilde{\beta}_\tau$ estimates from equation (1) and normalized by the mean as in equation (2). Panel (a) illustrates the response for pregnant women on welfare benefits and Panel (b) the response for children younger than age 13, whose mother is on welfare benefits. Shaded areas are 95% confidence bands.

# Appendix Tables

**Appendix Table 1: Rates for Welfare Benefits**

|  | DKK | USD |
|---|---:|---:|
| **Above age 25** | | |
| Without children | 9.857 | 1.450 |
| With children | 13.096 | 1.926 |
| | | |
| **Below age 25** | | |
| Not living with parents | 6.351 | 934 |
| Living with parents | 3.065 | 451 |
| With children | 13.096 | 1.926 |
| Pregnant after week 12 | 9.857 | 1.450 |

Notes: The rates are pre-tax. The USD exchange rate is USD 1 = DKK 6.8.

**Appendix Table 2: Distribution of Paydays on Weekdays**

|           | Number (N) | Percent (%) |
|-----------|------------|-------------|
| Monday    | 28         | 14          |
| Tuesday   | 27         | 14          |
| Wednesday | 30         | 15          |
| Thursday  | 30         | 15          |
| Friday    | 85         | 43          |
| Total     | 200        | 100         |

Notes: This table illustrates the distribution of the 200 paydays on weekdays. 43% of all paydays fall on a Friday. The period runs from March 18, 2020, to November 14, 2016.

**Appendix Table 3: Overview of Special Days Controlled for in Regressions**

| Special Day | Date |
|---|---|
| New Year's Day | January 1 |
| Palm Sunday | March/April |
| Maundry Thursday | March/April |
| Good Friday | March/April |
| Easter Sunday | March/April |
| Easter Monday | March/April |
| Prayer Day | April/May |
| May 1st | May 1 |
| Ascension Day | May/June |
| Day after Ascension Day | May/June |
| Whit Sunday | May/June |
| Whiit Monday | May/June |
| Constitution Day | June 5 |
| Christmas Eve | December 24 |
| Christmas Day | December 25 |
| Day after Christmas | December 26 |
| New Years Eve | December 31 |

Notes: This table provides the *special days* that I control for in regression equation (1) and (3).

**Appendix Table 4: Classification of Drugs for Chronic Conditions**

| ATC code | Active chemical agent |
|---|---|
| **Ace inhibitors** | |
| C09AA01 | captopril |
| C09AA02 | enalapril |
| C09AA03 | lisinopril |
| C09AA04 | perindopril |
| C09AA05 | ramipril |
| C09AA06 | quinapril |
| C09AA07 | benazepril |
| C09AA09 | fosinopril |
| C09AA10 | trandolapril |
| **Beta blockers** | |
| C07AA03 | pindolol |
| C07AA05 | propranolol |
| C07AA06 | timolol |
| C07AA07 | sotalol |
| C07AA16 | tertatolol |
| **Oral anti-diabetics** | |
| A10BA02 | metformin |
| A10BB01 | glibenclamide |
| A10BB03 | tolbutamide |
| A10BB07 | glipizide |
| A10BB09 | gliclazide |
| A10BB12 | glimepiride |
| A10BG03 | pioglitazone |
| A10BX02 | repaglinide |
| **Statins** | |
| C10AA01 | simvastatin |
| C10AA02 | lovastatin |
| C10AA03 | pravastatin |
| C10AA04 | fluvastatin |
| C10AA05 | atorvastatin |
| C10AA07 | rosuvastatin |
| **Birth Control Pills** | |
| G03AA | Progestogens and estrogens, fixed combinations |
| G03AB | Progestogens and estrogens, sequential preparations |
| G03AC | Progestogens |
| G03HB01 | cyproterone and estrogen |
| **Antipsychotics** | |
| N05A | |

Notes: I choose the drugs based on Skipper et al. (2017). Definition of drug groups comes from the Danish Health Data Agency (medstat.dk) and WHO Collaborating Centre for Drug Statistics Methodology (WHOCC).

**Appendix Table 5: Classification of Antibiotics**

| ATC code | Active Chemical Agent |
|---|---|
| **Antibiotics** | |
| J01 | Antibacterials for systemic use |
| **Antibiotics, narrow-spectrum** | |
| J01CE | Beta-lactamase sensitive penicillins |
| J01CF | Beta-lactamase resistant penicillins |
| J01EA | Trimethoprim and derivatives |
| J01EB | Short-acting sulfonamides |
| J01FA | Macrolides |
| J01FF | Lincosamides |
| J01XA | Glycopeptide antibacterials |
| J01XC | Steroid antibacterials |
| J01XD | Imidazole derivatives |
| J01XE | Nitrofuran derivatives |
| J01XX | Other antibacterials |
| **Antibiotics, broad-spectrum** | |
| J01AA | Tetracyclines |
| J01CA | Penicillins with extended spectrum |
| J01CR | Combinations of penicillins, incl. beta-lactamase inhibitors |
| J01DB | First-generation cephalosporins |
| J01DC | Second-generation cephalosporins |
| J01DD | Third-generation cephalosporins |
| J01DH | Carbapenems |
| J01EE | Combinations of sulfonamides and trimethoprim, incl. derivatives |

Notes: Definition of antibiotics, including division on broad and narrow-spectrum is from the Danish Health Data Agency (medstat.dk) and WHO Collaborating Centre for Drug Statistics Methodology (WHOCC).

**Appendix table 6: Indication for Antibiotic Prescriptions**

| Indication | N | % |
|---|---:|---:|
| against cystitis | 1,381,472 | 12.42 |
| against urinary tract infection | 1,216,465 | 10.93 |
| against skin and soft tissue infections | 1,077,646 | 9.69 |
| against pneumonia | 1,075,693 | 9.67 |
| against inflammation | 992,113 | 8.92 |
| against infection | 939,879 | 8.45 |
| against sore throat | 677,186 | 6.09 |
| against middle ear inflammation | 434,579 | 3.91 |
| against sinusitis | 424,656 | 3.82 |
| for the prevention of urinary tract infection | 301,265 | 2.71 |
| against acne | 192,404 | 1.73 |
| against Chlamydia / mycoplasma infection | 136,284 | 1.22 |
| against Borrelia infection | 93,061 | 0.84 |
| against bloating in chronic obstructive pulmonary disease (COPD) | 81,42 | 0.73 |
| against the erysipelas | 61,326 | 0.55 |
| acute exacerbation of chronic bronchitis | 55,792 | 0.50 |
| against stomach ulcers (eradication of Helicobacter pylori) | 53,739 | 0.48 |
| against diarrhea | 37,101 | 0.33 |
| against inflammation of the intestine | 27,338 | 0.25 |
| against infected eczema | 27,194 | 0.24 |
| against bronchitis | 26,441 | 0.24 |
| against staphylococci. incl, eradication of carrier state | 25,584 | 0.23 |
| against inflammation of the epididymis | 24,79 | 0.22 |
| against genital inflammation | 18,151 | 0.16 |
| against skin disorder | 17,373 | 0.16 |
| against impetigo | 16,868 | 0.15 |
| against bacterial infection in bones and joints | 15,229 | 0.14 |
| against inflammation of the urethra | 12,148 | 0.11 |
| against inflammation of the vagina | 11,456 | 0.10 |
| for the prevention of heart valve inflammation | 10,899 | 0.10 |
| for the prevention of malaria | 10,759 | 0.10 |
| against animal or human bites | 9,614 | 0.09 |
| against renal pelvic inflammation | 9,286 | 0.08 |
| against infection of the skin | 8,82 | 0.08 |
| against wound infection | 8,424 | 0.08 |
| against chronic urinary tract infection | 8,216 | 0.07 |
| on breast infection | 6,488 | 0.06 |
| for the prevention of severe infection | 6,446 | 0.06 |
| against scarlet fever | 5,395 | 0.05 |
| against severe infection | 3,936 | 0.04 |
| against whooping cough | 3,909 | 0.04 |
| against inflammation of the prostate gland | 2,991 | 0.03 |
| against rosacea | 2,199 | 0.02 |
| against gonorrhea | 1,918 | 0.02 |
| *no indication* | *1,559,288* | *14.01* |
| N total | 11,126,217 | 100.00 |

Notes: This table provides summary statistics on the indication (diagnosis) for antibiotic prescriptions for the full Danish population, 2015-2019.